# Support of Interactive 3D/4D Presentations by the Very First Ever Made Virtual Laboratories of Antennas

*Nikolitsa YANNOPOULOU[1], Petros ZIMOURTOPOULOS[2]*

[1] Antennas Research Group, Palaia Morsini, Xanthi, Thrace, Hellas, EU
[2] Antennas Research Group, Dept. of Electrical Engineering and Computer Engineering, Democritus University of Thrace, V. Sofias 12, 671 00 Xanthi, Thrace, Hellas, EU

yin@antennas.gr, pez@antennas.gr, www.antennas.gr

**Abstract.** *Based on the experience we have gained so far, as independent reviewers of Radioengineering journal, we thought that may be proved useful to publicly share with the interested author, especially the young one, some practical implementations of our ideas for the interactive representation of data using 3D/4D movement and animation, in an attempt to motivate and support her/him in the development of similar dynamic presentations, when s/he is looking for a way to locate the stronger aspects of her/his research results in order to prepare a clear, most appropriate for publication, static presentation figure. For this purpose, we selected to demonstrate a number of presentations, from the simplest to the most complicated, concerning well-known antenna issues with rather hard to imagine details, as it happens perhaps in cases involving Spherical Coordinates and Polarization, which we created to enrich the very first ever made Virtual Laboratories of Antennas, that we distribute over the Open Internet through our website Virtual Antennas. These presentations were developed in a general way, without using antenna simulators, to handle output text and image data from third-party CAS Computer Algebra Systems, such as the Mathematica commercial software we use or the Maxima FLOSS we track its evolution.*

## Keywords

Antennas, virtual laboratories, dynamic presentation, interactive presentation.

## 1. Introduction

The very first ever made Virtual Laboratories of Antennas, which are available on the Internet through Virtual Antennas website [1], are exclusively based on our alternative form of Antenna Theory [2], and are founded, by applying the learning by teaching method to Antennas education [3], in order to support the quick comprehension of 3D space matters unavoidably related to antennas. The initial material of Virtual Antennas was prepared during the years 1996-1997 and presented to Antenna students during the spring semester of 1998 [4]. On 1999, the existence of this visual material was announced to the first ever appeared on the web EMLIB Electromagnetics Library, maintained those years by Jet Propulsion Laboratory of NASA [5]. On the same year, all of the material was approved for inclusion to MathSource repository of Mathematica [6]. Since 2000 and until today, the Virtual Antennas material is constantly increased and improved. Moreover, a number of websites either use this material or suggest its use [7]. On 2009, a sample of its current development state was approved for publication by the Wolfram Demonstrations Project [8], an event that was announced in our Creative Commons Network web pages [9]. During the last years, our voluntary reviewing work revealed some needs of potential authors -especially the young ones- related to the presentation of their research results, which perhaps may be confronted by similar to our interactive presentation techniques, and thus we decided to present in this paper our continuing work for the Virtual Laboratories of Antennas.

## 2. Presentation Development and Use

The implementation of a presentation idea demands the expression of the theoretical idea formulation in the CAS language we select to produce output data in the appropriate file format for movement or/and animation (we would like to emphasize that in the case of Mathematica we use, while we abandoned the package written by Novak [10], we still run the one written by Donley [11]). These files are: (a) WRL (WoRLd: Open by Web3D Consortium), a plain text file with a known structure described in the VRML Virtual Reality Modeling Language [12] for a visual object, (b) a bunch of images in a selected non-destructive format; usually BMP, and (c) NBP text file (Note Book Player: Mathematica proprietary). After that, the development of the presentation requires: (a) either a text editor to correct bugs or to add special features into WRL output file, (b) a graphics editor to align images or correct blemishes, as well as, a video editor to handle these images as frames of the final movie file in a selected format, usually AVI (Audio Visual Interleave: Microsoft proprietary). It is worth to notice that according to our experience, the non-computing time needed: (a) to design the idea implementation, has been reduced by the years, from a couple of weeks at the beginning to just a couple of days now, and (b) to develop the presentation, after a tedious, routine work of editing each frame, varies considerably according to the frame theme, e.g. from 1.5 minutes for that in Fig. 7, to 10 minutes of Fig. 10.

The software needed for presentations to work under MS Windows is: a web browser (MS-IE 3.02r+), a VRML viewer add-on to that browser (WorldView 2.1), an AVI player (Mplayer2), and the Mathematica 7 Notebook Player.

## 3. 3D/4D Presentation Samples

In order to abbreviate the figure descriptions in the following, we have to notice that: (a) for every animation presentation, a typical full-window screen capture of the recommended AVI player is shown after its pause button was pressed, (b) for every movement presentation, a typical full-window screen capture of the recommended online WRL viewer, with a cursor perhaps to indicate the existence of an additional feature, and (c) for both presentation types, the basic colors (R, G, B) are used in that order to correspond either to the CCS Cartesian Coordinate System (x, y, z) axes, (xoy, yoz, zox) planes, and its unit vectors, or to the SCS Spherical Coordinate System (r-radius, theta-semi-circle, phi-circle) curves, (r-sphere, theta-semi-cone, phi-semi-plane) surfaces, and its unit vectors, while the same color correspondence holds for the radiation pattern cuts by the mentioned CCS planes or SCS surfaces.

After that, short descriptions of the samples, are following, while all of these presentations will be always available in authors' group repositories in Virtual Antennas [13] and GoogleCode [14].

Fig. 1-left shows 1 frame out of 12 of an AVI repeated-for-ever animation for the 3D plane-time presentation of the considered as time-harmonic sinusoidal current wave amplitude, on a 3 wavelengths portion of a long thin wire loop antenna terminated on some complex impedance. The propagated current wave p is decomposed in the dual couples: (incident wave i, reflected wave r) and (standing wave s, transmitted wave t), while the imposed letters and arrows on the figure indicate these current waves as well as the direction of their motion.

Fig. 1-right shows a screen-shot of a WRL presentation with 5 additional predefined view points of 12 frames animation, as the cursor indicates, of a 4D space-time asynchronous movement, from a view-point in the 1st-octant. In essence, this presentation is an alternative geometric meaning of the considered as standing wave current on an adequate length of a thin wire dipole antenna.

Fig. 2-right shows a screen-shot of a WRL presentation of a 4D space-time asynchronous movement from a view-point in the 1st-octant, for the triples of SCS unit vectors in the shown directions, that is 13 triples on the CCS main-planes, as well as, 2 undefined triples on the irregular z-axis of the CCS-to-SCS transformation.

Fig. 2-left shows a screen-shot of a WRL presentation of 4D space-time asynchronous movement from a view point in the 1st-octant, of the 3D difference volume element formed by adjacent SCS coordinate surfaces and lying between the 1st and 2nd octant.

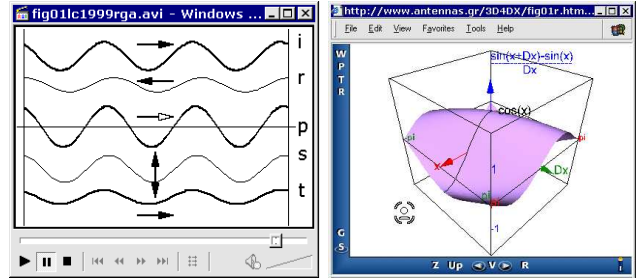

**Fig. 1.** Current waves - Standing current-voltage waves relation.

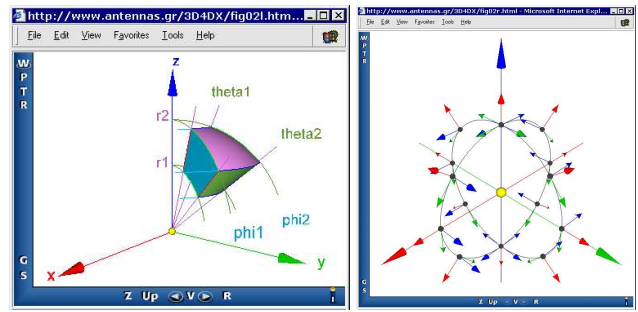

**Fig. 2.** SCS: Volume element - Unit vectors.

Fig. 3-left shows a screen-shot of a WRL presentation of a 4D space-time asynchronous movement from a view-point in the 1st-octant, for all 3 SCS surfaces, curves, and unit vectors in a direction of the 1st- octant.

Fig. 3-right, shows 1 frame out of 37 of an AVI animation for the 4D plane-time presentation of a SCS phi-circle curve as the intersection between 2 SCS surfaces, that is of a r-sphere and a theta-semi-cone, while theta angle -which changes from 0° to 180°- has the value of 50°. Different colors illustrate the inner and outer surfaces of the r-sphere and a cut off spherical section between adjacent meridians permit us to see the interior of that sphere.

Fig. 4-left shows a screen-shot of a WRL presentation of a 4D space-time asynchronous movement for three different polarizations at three different points: linear, circular and elliptical on each of the coordinate axes x, y and z respectively to emphasize that, in general, the polarization of an antenna is not constant.

Fig. 4-right shows 1 frame out of 20 of an AVI animation for the 4D space-time presentation of the non-linearly CCW polarized time-harmonic real electric field *E*, from an antenna, as well as, its decomposition to two

linearly polarized time-harmonic real electromagnetic fields, as they are uniquely defined by the constant of time linearly independent space vectors $E_c$ and $E_s$.

Fig. 5-left shows the last frame out of 23 of an AVI animation for the 4D space-time presentation of the path traced by the electric far-field arrow tip, which is CW elliptically polarized in the direction of y-axis.

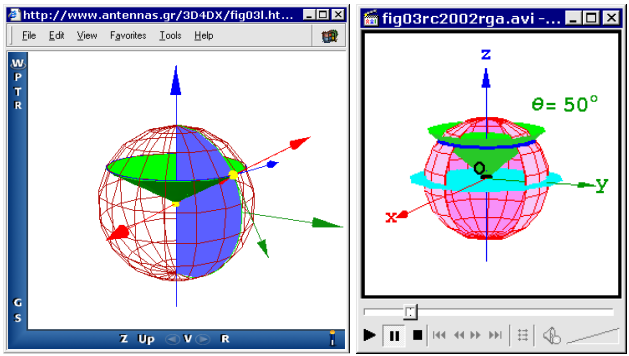

**Fig. 3.** SCS: Coordinate curves, surfaces and unit vectors.

Fig. 5-right shows 1 frame out of 19 of an AVI for the 4D space-time presentation of a normalized CCW elliptically polarized electromagnetic far-field in the shown direction of propagation, from a point of view in the 1st octant of an observer mirrored to (r, theta) plane.

Fig. 6 shows 3 frames out of 73 of an AVI for the 4D space-time presentation of the direction dependency of antenna polarization from a center fed short crossed dipoles on y and z axes, which have a current phase difference of 90°. The three different polarizations are shown with their orientation for an observer on the xOy plane (theta = 90°) and in three different phi angles of 0°, 240° and 270°.

Fig. 7 shows 4 frames out 38 of an AVI animation for the 4D space-time presentation of the space radiation pattern and its three main-planes cuts, for the specific case of an antenna consisting of 2 dipoles, with the same direction of the unit vector (0.2, 0.4 0.894), each of 2.4λ length long, with their centers placed 0.25λ apart on an axis with a unit directional vector (0.3 0.5 0.812), fed with a 30° current phase difference.

Fig. 8 shows a screen-shot of a WRL presentation with 1 additional predefined view point of 12 frames animation possibility, as the cursor indicates, of a 4D space-time asynchronous movement, from a view-point in the 1st-octant. This presentation concerns the interior of an anechoic chamber and imitates a discone antenna rotation in order to visualize the way by which the measurement of a plane-cut of its radiation pattern is accomplished.

Fig. 9 shows a sophisticated combination of a movement with an animation presentation, which in fact is an application that was developed using the version 7 of Mathematica, in a way that permits the definition of any (theta, phi) direction in space for a dipole of variable length. Notably, this application has all the available VRML movement features, with 6 predefined view points, while 3 different animations can be run simultaneously. In addition, the mesh, opacity, rendering goal, and the evaluation step for the radiation pattern can be also defined.

Fig. 10 shows the last frame out of 100 of an AVI for the 4D space-time presentation of a detailed study for some of the linear dipole characteristics regarding the plane and space radiation patterns, the direction from the dipole axis of their first common maximum, the directivity, and the input radiation resistance; this really is a 4-tuple of frames.

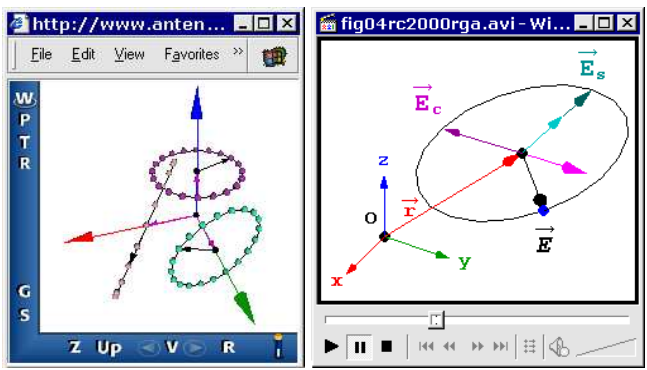

**Fig. 4.** Polarization: Directional dependence - Non-linear CCW.

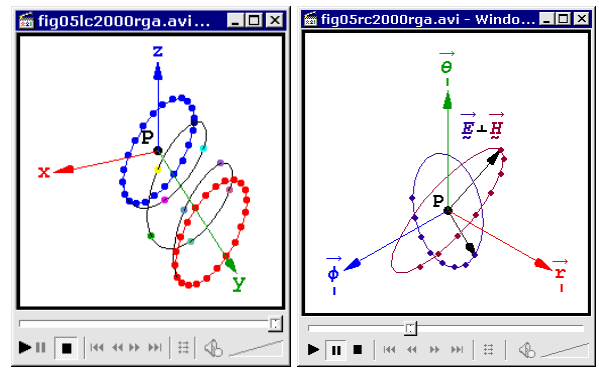

**Fig. 5.** Polarization: CW elliptical - Far-Field CCW elliptical.

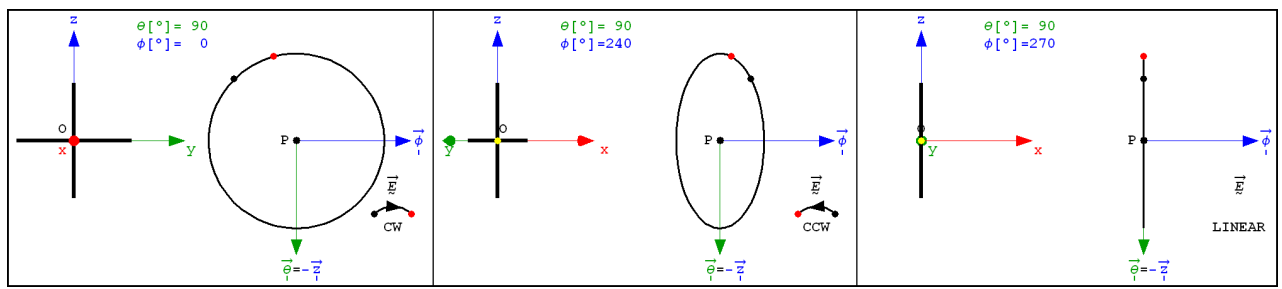

**Fig. 6.** Polarization: CW Circular - CCW Elliptical - Linear.

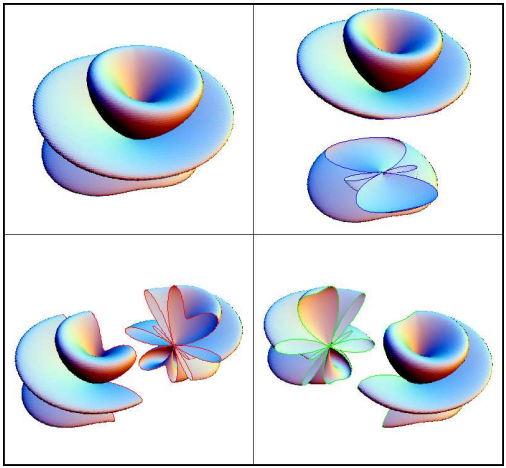

**Fig. 7.** Radiation Pattern: 3D and its 2D main-plane cuts.

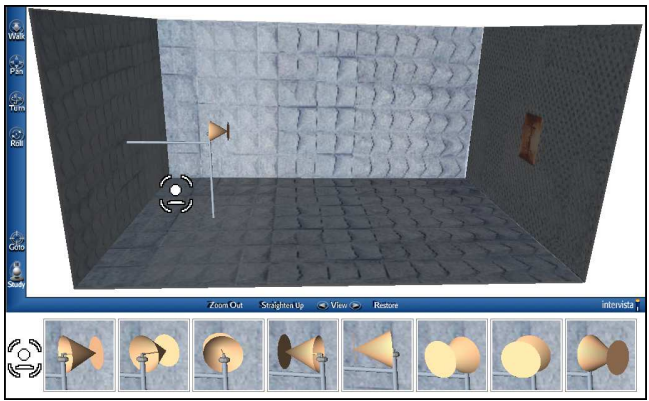

**Fig. 8.** Anechoic chamber - discone antenna.

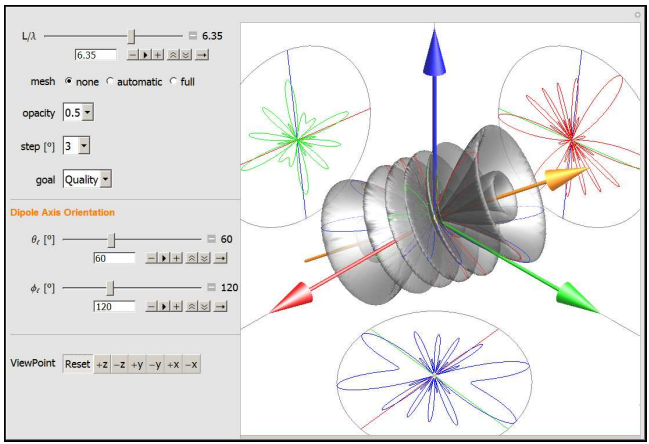

**Fig. 9.** Radiation Pattern: A composite 3D/4D presentation.

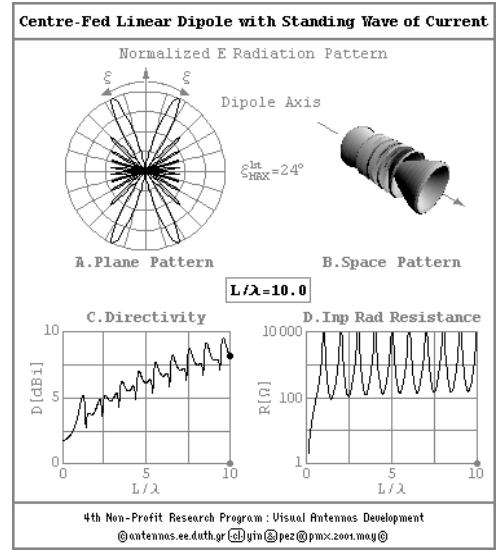

**Fig. 10**. Antenna characteristics.